# Emergent Spatial Characteristics from Strategic Games Simulated on Random and Real Networks

Louis Zhao[a,1], Chen Ye Gan[a,1,2], and Minglu Zhao[b,1]

[a]Department of Mathematics, University of California, Los Angeles; [b]Department of Statistics, University of California, Los Angeles

**Complex networks are a great tool for simulating the outcomes of different strategies used within the iterated prisoners' dilemma game. However, because the strategies themselves rely on the connection between nodes, then initial network structure should have an impact on the progression of the game. By defining each interaction in terms of a prisoner's dilemma and using its payoff matrix as a basis for investigation, we implemented players with various interaction and edge attachment strategies, and ran this dynamic process on real and random networks with varying network structure. We found that, both network size and small world properties played an important role in not only deciding the convergence rate of the simulation but also the dominant status of nodes, under the conditions where identical strategies are employed by every player.**

Iterated Prisoner's Dilemma | Simulation | Complex Network

P<small>RISONER</small>'s dilemma is a paradox where two parties competing to maximize self interest do not result in the optimal output for both. While a stable Nash equilibrium where both parties defect is reached in a single round situation, iterated versions are easily perturbed by initial conditions as well as probabilistic or conditional decisions made during each iteration.

One such simulation to model various routes of evolution is the Evolution of Trust (1) — an online game where simulations of single shot and iterated prisoner's dilemma are shown. This game features an array of players, which includes, but is not limited to, the copycat, which copies whatever move the opponent made the previous iteration, the always cooperate, the always defect, and grudges, who cooperates until the opponent defects for the first time, after which the grudges always defects, and the detective, that begins with a set number of cooperates and defects and determines from there if a copycat strategy or an always cheat strategy should be employed. In their simulations, in a world with equal number of copycats, always defects, and always cooperates, copycat dominates. If filled with equal parts copycat, grudger, and detective, copycat still wins. However, if the payoff matrix was altered in favor of defecting, or if the proportion of players changed, the outcome could be a different dominant player type, or an infinite oscillation between two player types.

However, that model only presented game simulations in which the entire network was completely connected, such that during every iteration, every player would interact with every other player. In real world models, however, that is not the case. Barabási-Albert (BA) Model (2) and Watts-Strogatz (WS) models (3) that we used were more representative of the actual clustering of nodes in real life network. Also, the simulation presented above used nodes that had different strategies which was effective in presenting the dominant player type. We, on the other hand, also want to investigate into the long run convergence of players with similar strategies, varying only the network structure, and observe how initial network structures impacted emergent dominance of a binary node type (i.e. always cooperating or always defecting), rather than players whose strategies would never change and would eventually die out similar to natural evolution.

Previous literature have looked into similar processes of analyzing the effects of the iterated prisoner's dilemma game on network structure. When edge reformation and semi-omnipotent strategies are incorporated into simulations on Erdős-Renyi (4) networks, overall cooperation density increases (5). Initial network structure also plays a role in the progression of an iterated prisoner's dilemma game. Cooperation strategies converge faster to a Nash Equilibrium when the initial network has a higher clustering coefficient and lower geodesic distance (6). There has however not been a substantial amount of previous work done on comparing the iterated prisoner's dilemma game on randomly generated models to real life social networks which is the focus of this paper.

When constructing strategies for our models, we examined many popular strategies, including the tit-for-tat strategy that has commonly won in many different simulation that used the prisoner's dilemmas as a basis. We also took note of the relatively novel zero-determinant strategies proposed by Press and Dyson (2012) in which players can unilaterally force a linear payoff relationship between themselves and an opponent (7). We also considered players with m-level thinking, where nodes are able deduce using the logic of "I know that you know that I know", for N rounds of recursion (8). We also looked at strategies that were effective in evading a certain Markov equilibrium, by playing within a certain timescale (9). In the end, we chose to use a few simpler strategies, partly because the some of the math presented were difficult to translate into code under the constraints of our simulation, and partly

> **Significance Statement**
>
> Games on networks play a crucial role in the field of behavioral economics and game theory. By defining each interaction to be a prisoner's dilemma game, we examined how different strategies and their inevitable outcomes can be affected by changing the network structure on which the game is played. In this paper, we tested different strategies using different models of network generation along with various real life networks and found high correlation between local clustering and the dominance of cooperators. It gives insight into the types of structure on which cooperation emerges.

because we wanted more predicable strategies for generation, allowing us to distinguish between the influence of the choice of strategy and the choice of network structure on which the simulation is played upon.

## Methodology

Our simulation is broken up into four large components. First, we generate network models by using random networks using ER Models, WS Model, and BA model, and culled data from real networks. The basis of our network choices are described in the network subsection. Next, we ran simulations using predefined rules for interactions during each iteration as well as evolved the network with dynamic attachment and edge removal rules. Using network evolution data from the previous step, we calculated various centrality measures as well as performed community defection on each iteration. Finally, combining raw, generated, and calculated data, visualizations in the form of network graphs, line plots, scatter plots, bubble plots, histograms, and animations, were generated and used to analyze and make conclusions about the properties of the evolution of each initial network given the specific parameters and strategies for interaction during each iteration.

**Real Social Networks.** We tested our simulations on two real social networks of varying sizes and structure. The first social network is Zachary's Karate Club, first described by Wayne W. Zachary in his 1977 paper *An Information Flow Model for Conflict and Fission in Small Groups*(10). It has 34 total nodes and 78 total edges and is a textbook example of a social network that displays community structure. The second social network used within this project is a randomly generated sub graph of a network of Facebook users connected by their friendship status on Facebook(11). The original network categorized the users into different social circles based on shared interests and common attributes. To construct our network, we took a random sample of these social circles and combined their nodes and edges into a new network. We did this because we did not have enough computing power to run all our simulations and analysis on the original network. The resulting network had 1074 nodes and 31863 edges.

**Network Models.** We used three different models to generate the networks on which we tested our simulation of the iterated prisoner's dilemma game. The three models were the probabilistic Erdõs-Renyi Model, Watts Strogatz Model, and the Barabási-Albert Model. In the probabilistic Erdõs-Renyi Model, a network $G(n,p)$ is constructed such that every edge within the network is included with probability $p$ independent from every other edge. $G(n,p)$ is then the ensemble of all simple networks with $n$ nodes in which each network $G$ appears with probability

$$P(G) = p^m(1-p)^{(\binom{n}{2}-m)} \quad [1]$$

where $m$ is the number of edges in the network. (4) Because we are examining the effects of network structure on the prisoner's dilemma game, we believed that a random network served as a great baseline because of its simplistic structure and lack of small world properties.

In the Watts Strogatz Model, we first define a network with a regular lattice structure on $n$ nodes. The nodes are connected to the nearest $k$ even nodes nearest to it. We then rewire each edge with probability $p$ to a random node chosen uniformly from the network to form our final network(3). Because of the existence of shortcuts within the Watts Strogatz Model, it has a much higher clustering coefficient and much shorter average path length than the ER model. This gives the network small world properties and thus makes it a much closer representation of real life social networks than the ER model. However its degree distribution does not follow a power law distribution but rather that of a random graphs (12).

The Barabási-Albert Model, on the other hand, does follow a power law degree distribution by utilizing two elements that incorporate both the growing characteristic of the network and preferential attachment. The model first starts with a small number ($m_0$) of vertices. Every step, we add a new vertex with

$$m \leq m_0$$

edges that link the vertex to $m$ different vertices within the network. While adding the edges, preferential attachment is incorporated by assigning a probability $p_i$ of attachment to each existing node that is proportional to the existing node's current degree $k_i$.(2) The probability $p_i$ that the new node is connected to node $i$ with degree $k_i$ is :

$$p_i = \frac{k_i}{\sum_j k_j} \quad [2]$$

Using these three network models, we generated 7 different undirected unweighted simple networks to run our simulation on. These networks are broken down into different categories based on the total number of nodes and the specific model used to generate them. For the Erdõs-Renyi model, we generated two 34 node networks with varying probability values ( $p = 0.1$ and $p = 0.9$) and one 1000 node network using $p$ value 0.07 to match the number of edges found in the Facebook network. By changing the probability of which edges will form between a given pair of nodes, we can better understand how our simulation might differ by varying number of edges and thus varying network densities. For the Watts Strogatz and Barabási-Albert Model, we generated two networks each of two different sizes (34 total nodes and 1000 total nodes). These different sizes are meant to represent our real life networks which has 34 and 946 nodes respectively and allows us to keep constant network size while focusing our analysis on the actual differences caused in network structure by the different models. Our parameters for constructing the Barabási-Albert and Watts Strogatz models were also picked in such a way that we maintained the same number of edges as the real social networks. For example, because the Facebook network had an average degree of around 30, we picked large enough parameters ($k = 60$ and $m = 32$) such that our Watts Strogatz and BA networks also had an average degree of around 30. This was to remove any inconsistencies and unwanted effects that might have been caused by having networks with differing number of edges. Lastly, when generating our Watts Strogatz network, we chose a probability of shortcut $p$ of 0.01 such that we can maintain a fairly low geodesic distance while also having a high clustering coefficient, thus preserving the small world properties of the Watts Strogatz network.

**Simulation.** In this paper, we introduce a set of strategies for a repeated-interaction game based on the famous prisoner's dilemma problem. Starting with an initial network with N nodes and initial adjacency matrix $A_0$, each node $i \in \{1, 2 \ldots N\}$ has an initial status $S_{i0}$ being either a cooperator ($S_{i0} = 0$) or a defector ($S_{i0} = 1$). The status of the nodes are randomly assigned, with a $r_0$ fraction of the nodes being cooperators $0 < r_0 < 1$. The nodes can switch status based on the specific strategy used in the game setting after each simulation round. During each simulation round $t \in \{1, 2 \ldots K\}$, $K \in \mathbb{R}$ is the total number of iterations, and two nodes interact if they are neighbors in the network ($A_t[i, j] = 1$). In the current setting, cooperators always cooperate, and defectors always defect, such that there are no 'accidents' or 'miscommunications'. Each node $i$ will receive a reward that depends on both its current status and the status of its neighbor $j$ with whom it just interacted, corresponding to the predefined payoff matrix $M$, $P_{itj} = M[S_{it}, S_{jt}]$. The total payoff of node $i$ at round $t$, denoted by $P_{it}$, is defined as $P_{it} = \sum_{j \in \{1,2\ldots N\}, A_t[i,j]=1} P_{itj}$. After each round, edges connecting two nodes $i$ and $j$ will be broken if any one of them is a defector: $A_{t+1}[i,j] = \begin{cases} 1 & \text{if } S_{it} + S_{jt} = 0 \\ 0 & \text{otherwise} \end{cases}$. New edges will be formed based on the simulation strategy being used in the current simulation episode. All networks (starting networks and simulated networks) are simple networks that are undirected, unweighted, and have no self-edges or multi-edges.

Each of the simulation methods that we use here can be considered as having two components: 1) the strategy to change the nodes' status, and 2) the strategy to form new edges between nodes. The process of changing status can be interpreted as agents switching sides in the game. In this paper we present two strategies for nodes to change status: 1) simulating self-focus agents who do not care about neighbors' behavior and outcome, and 2) simulating conforming agents whose status highly depend on neighbors' status. For a self-focus node $i$ at simulation round $t$, it changes status (switches from cooperator to defector, or from defector to cooperator) whenever $P_{it} < P_{i(t-1)}$, meaning that the total payoff of node $i$ from the current simulation step $t$ is smaller than last step. For node i in the conformity status-changing strategy, it switches status if $P_{it} < \frac{\sum_{j \in \{1,2\ldots N\}, A_t[i,j]=1} P_{tj}}{\sum_{j=1}^{N} A_t[i,j]}$. In this case, a node decides to change sides if its payoff from the current time step is smaller than the average of their neighbors' payoff from the same simulation step. However, here the node changes status in a more sophisticated way. Specifically, $S_{i(t+1)} \begin{cases} 1 & \text{if } \sum_{j=1}^{N} A_t[i,j] * S_{jt} > \frac{1}{2} \sum_{j=1}^{N} A_t[i,j] \\ 0 & \text{otherwise} \end{cases}$. In this way, node $i$ will switch to the side taken by a larger number of its neighbors, and we can also say that it chooses to "conform" to its neighbors.

The second major component of the simulation is the edge-forming process, which has two implications: 1) The process of forming new edges simulates the process of looking for new friends in a network setting; 2) New edges are formed only when there are broken edges in the most recent round, a process that helps maintain the number of edges in the network to some extent. In the game setting, if an edge between two nodes is broken due to the interaction between one cooperator and one defector, then the cooperator will form a new edge. If an edge between two defectors is broken, then we randomly sample one of the two defecting nodes to form a new edge. During the interactions, each node keeps a record of the number of new edges it should form at round $t$, denoted by $e_{it}$ for node $i$. Here, we introduce two strategies to form new edges which correspond to two different kinds of nodes in the simulation. Intuitively, a node can form new edges based on 1) social punishment, or grudge, towards defectors that it interacted with or 2) social reputation of potential nodes. In the first scenario, each node $i$ keeps a record of the status of the nodes that it interacted with and avoids forming edges with those that were defectors in such interactions for $g$ steps. In other words, the nodes "hold a grudge" towards the defectors they came across for $g \in \mathbb{R}$ rounds. Each node $i$ will then randomly select $e_{it}$ nodes that are not on the "blacklist" and not its neighbor (to avoid multi-edges) to connect with. If $e_{it}$ is larger than the number of qualified nodes, node $i$ will then form edges with all nodes that fulfill the requirement. In the second case, new edges are formed based on the "cooperation history" of nodes. Nodes are more likely to form edges with nodes that have a higher ratio of cooperation in the history of its status. Specifically, at each round $t$, each node would calculate the defecting score of all other nodes. Defecting score of node $i$ is defined as $r_{it} = \frac{1}{t} \sum_{l=1}^{t} S_{il}$, with a higher defecting score showing a higher percentage of being a defector in the interaction history. When selecting nodes to form edges with, the list of potential new neighbors of node $i$ includes all nodes except $i$ itself and neighbors of $i$. Node $i$ would calculate the defecting score of the qualified nodes, randomly shuffle the nodes, and then select each node j with probability $q_j = max(1 - r_{jt}, Q_0)$ until $e_{it}$ nodes have been selected. $Q_0 \in \mathbb{R}$ is a constant used to account for the initial burn-in period of the defecting score calculation to ensure some level of exploration in selecting nodes.

**Table 1. Five simulation strategies**

| Strategy ID | Status-changing | Edge-forming |
|---|---|---|
| 0 | Never change | No new edges |
| 1 | Self-focus | Social punishment $g = 1$ |
| 2 | Conformity | Social punishment $g = 5$ |
| 3 | Self-focus | Reputation |
| 4 | Conformity | Reputation |

In this paper, we report simulation results of five different simulation strategies. One of them is a base policy where the nodes never change status, and no new edges are formed. In the base policy, edges break with probability $m_0, 0 < m_0 < 1$, if any of the two interacting nodes is a defector. The other four policies incorporate combinations of the aforementioned simulation methods for the status-changing and edge-forming processes. Detailed information on the five strategies is specified in Table 1.

**Analysis and Visualization.** To visualize our data, several graph models were generated. Using the networkx package, a simple graph plot, with nodes distributed evenly Fruchterman-Reingold's algorithm (13), was generated, with node positions fixed for future iterations. This fixture of positions allowed edge forming and breaking pattern to be apparent over several iterations once graphs were strung together and animated.

Note, in the graphs, navy blue nodes are cooperators and deep crimson nodes are defectors, which correspond to the properties specified in each node's status property, as described in the simulation section. The colors of the edges, being teal, salmon, and orange, correspond to connections between cooperators, connection between defectors, and edges with one stub attached to a cooperator and the other connected to a defector, respectively.

For further visualizing of specific characteristics present in the network plot, we generated two accompanying animated histograms in sync with the generation rate of the network plot, one for node distribution (proportion of navy to crimsons nodes), and the other for edge distribution (teal, salmon, orange). This data is also coalesced into two line plots, one showing the evolution of nodes and the other of edges over time, with identical color schemes.

To visualize our centrality measure, we have compiled all the 4 different centrality data for each iteration (degree, eigenvector, katz, pagerank) onto a single line plot where each line of the same color represents one centrality measure over iteration counts. The data is presented on a scale of 0-1 by running Min-Max Normalization (14) on each point. The shape and lines types for each line plot is used to provide distinction between the various measures. In addition, measures are separated into averages for cooperating nodes and that of defecting nodes, distinguishable by their characteristic navy blue and crimson.

Further plots included scatter plots for every simulation showing the relationship between per round payoff and a node's probability of defection as measured by averaging status over previous iterations. This is again normalized onto a scale of [0, 1] on the x-axis.

Finally, in order to properly present data coming from our community defection algorithm, a greedy algorithm that categorizes nodes into separate communities based on increasing modularity (15), which included information of not just communities over time, the proportion of cooperators and defectors in every community, but also the number of communicates and the magnitude of each that was formed during iterations. Thus we utilized a bubble plot, where bubbles with the same color and present on the same vertical axis showed communities that formed within the same iteration. Their position on the y-axis showed the average status of the community (mostly defectors or cooperators) and the sizes of each bubble represented the literal size of the each community, only scaled up by a factor of 100 for visualization purposes.

### Results

**Base Case.** Results simulated using the base policy would serve as a baseline for analysis. By only breaking edges and never changing status or making new edges, the base policy produced a node distribution shown in Figure 1a, a constant proportion of cooperating to defecting nodes. The edge distribution graph (Figure 1b) shows an exponential decay of defector-defector and mixed edges in contrast with a logarithmic growth of cooperator-cooperator edges.

**Generated Networks.** The four figures in Figure 2 show how the proportion of nodes change over time for 3 different networks, the BA being 1000 nodes and the other 3 (ER, WS, and Karate

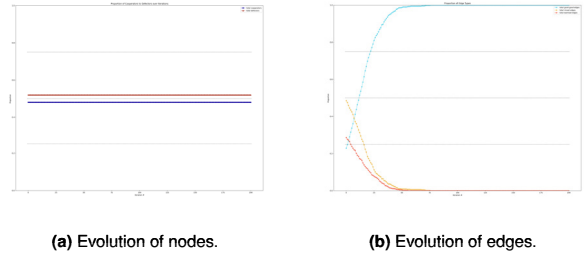

**(a)** Evolution of nodes.　　**(b)** Evolution of edges.

**Fig. 1.** Result under strategy 0 using a BA network with 34 nodes.

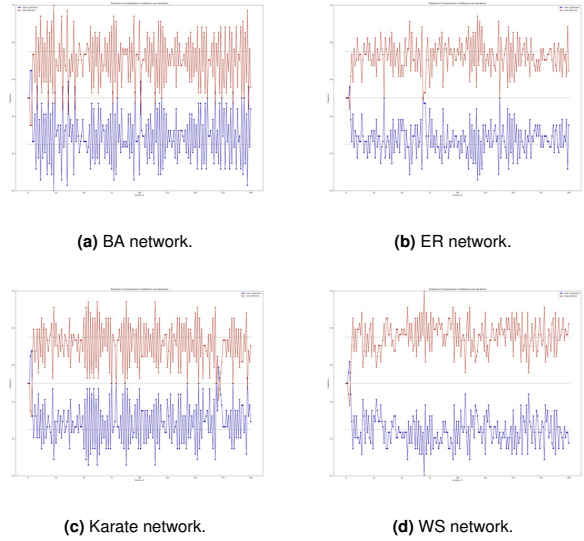

**(a)** BA network.　　**(b)** ER network.

**(c)** Karate network.　　**(d)** WS network.

**Fig. 2.** Evolution of nodes graph for results under strategy 1.

network) being 34 nodes each, all of which used strategy 1. In all the figures, despite oscillation, all evolution began showing the dominance of defectors, which their numbers, on average, reach much higher proportions than cooperators. In fact, while the amplitude of oscillations varied from graph to graph, the overall trend and proportion of approximate 0.75 to 0.25 defectors to cooperators holds for each of the network using strategy 1.

Of the models using strategy 3, the ER (Figure 3a) and karate (3b) networks exhibit logarithmic growth of defecting nodes, similar to that of our base strategy but at a much slower rate of convergence.

The BA(Figure 4a) and WS (Figure 4b) graphs, despite using the same strategy, tend to favor cooperators in the longer run as evident by the large clustering of cooperators and the isolation of defections in their respective bubble graphs.

**Facebook Network.** We then compared our simulation on the Facebook network against the same simulations on networks of equal sizes generated by Erdős-Renyi, Watts Strogatz, and Barabási-Albert models across 100 iterations. Figure 5 shows the various centrality measures for these simulations using strategy 3. The Facebook model is characterized by a high

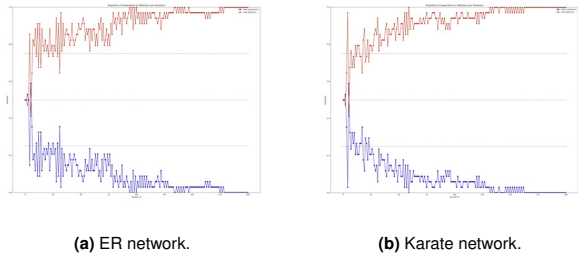

**(a)** ER network.    **(b)** Karate network.

**Fig. 3.** Evolution of nodes graphs for results under strategy 3 with 34 nodes.

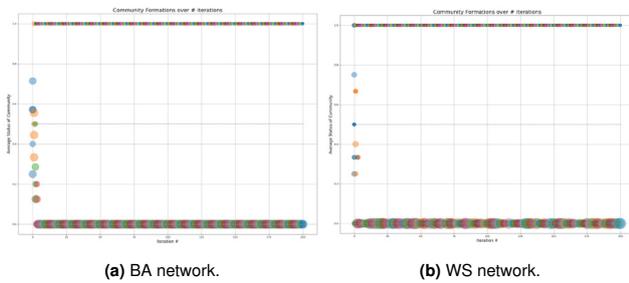

**(a)** BA network.    **(b)** WS network.

**Fig. 4.** Community detection plots for results under strategy 3 with 34 nodes.

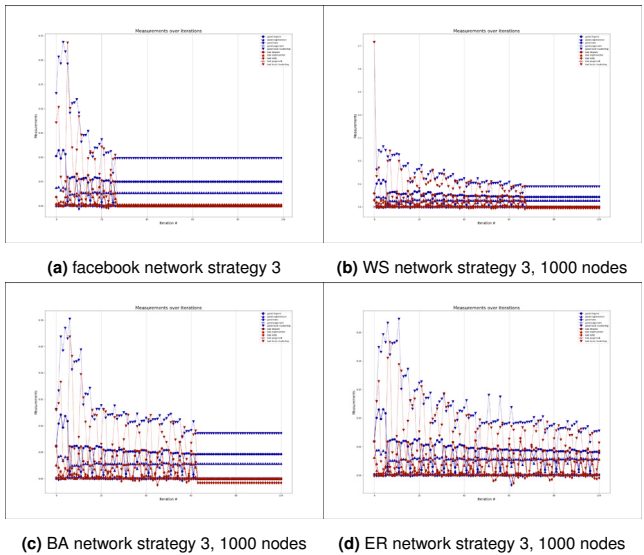

**(a)** facebook network strategy 3    **(b)** WS network strategy 3, 1000 nodes

**(c)** BA network strategy 3, 1000 nodes    **(d)** ER network strategy 3, 1000 nodes

**Fig. 5.** Centrality Analysis on Networks of size 1000

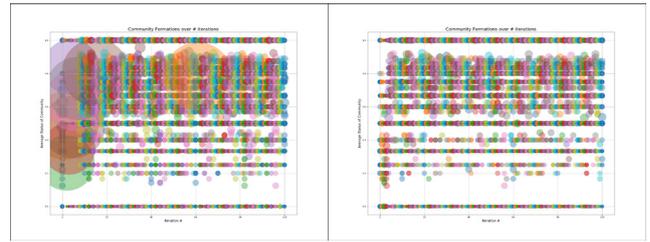

**(a)** BA network, low density, 1000 nodes **(b)** WS network, low density, 1000 nodes, community structure    community structure

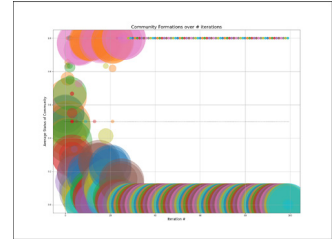

**(c)** Facebook network, community structure initial clustering coefficient of .619[*] and a power law degree distribution. The WS network also has a high initial clustering coefficient of .717 but not a power law degree distribution while the BA network has a power law degree distribution but a low clustering coefficient of .132. The ER graph, on the other hand, does not share any of these properties as it is randomly generated. When we ran our simulation with strategy 3 on these networks, we discovered that the Facebook network converged into a cooperator dominant network the quickest in 28 iterations, followed by the BA Model in 63 iterations, WS Model in 68 iterations, while the ER Model never converged within our allotted 100 iterations. Also, every simulation had a similar general structure across successive iterations where centrality measures decreased until reaching convergence. After converging, all three networks had a cooperator local clustering value of approximately .09 with a degree centrality value of approximately .05.

**Fig. 6.** Community Structure on Low density BA and WS networks vs. Facebook Network

When we ran the same models on parameters that resulted in much fewer edges: (Watts Strogatz: $k = 4$, p =0.01 and Barabási-Albert : $m = 3$), the networks failed to converge into a cooperator dominant state. Instead a fluctuating pattern emerged within the network where the total proportion between cooperators and defectors changed every iteration with the number of defectors always being greater than the number of cooperators. This can be seen on the community structure graphs shown in Figure 6. Here, the WS and BA networks have almost random community structures spread out across both cooperators and defectors and never converges while the Facebook network converges into a cooperator dominant state.

## Discussion

**Base Case.** The constant proportion of node types is trivially due to the fact that this policy defines no status changing conditions. The edge distribution stems from the fact that only edges are broken when one player is at a disadvantage during an interaction. By that logic, edges with one stub connected to a defector will have a chance of breaking every round. Since the number of edges broken is proportional to the total number of edges that connect to defectors, the rate of breaking slowly decays as the total number of said edges dwindles down to zero, resulting in an exponential decay that tapers out around 0. The remaining edges are simply the complement of the defecting edges, thus following a logarithm.

---

[*] for display purposes we removed the first iteration

**Effect of Blindness.** Strategy 1 assumes blindness to surrounding neighbors, which entails that by only comparing to a node's individual immediate last payoff, that node is essentially playing a single player version of prisoner's dilemma. Since our simulation assume homogeneity of node characteristics, all other nodes around are all interacting in the same fashion as single round prisoner's dilemmas with no memory of opponent's past status. In that situation, the best strategy, as pointed to by previous literature, is defecting. Thus every network, when using strategy 1, ends up with mostly defectors.

We assume, if the initial graph structure was untouched (edges weren't broken and formed every iteration), then this network would very quickly converge to all defectors

**Cooperator Convergence Analysis.** Strategy 3 was the only strategy that ended in a cooperator dominant convergence. However, this was only observed when applying the strategy to networks of larger size. On smaller networks, our results were very inconsistent where some networks favored cooperation while others favored defection. We attribute these inconsistencies to the small sizes of the networks themselves. Because the networks were so small, there is a lot of randomness that could have affected our outcomes when both generating the initial conditions for the networks and also during the simulation. Therefore, we found that we couldn't apply the conclusions made from our larger networks onto the results gathered from the smaller networks.

However, our results and findings were much more consistent when we increased the network size to 1000 nodes and increased the number of edges to 30000. When analyzing the simulation on 1000 nodes with strategy 3, all of our networks except for the ER network converged into a cooperator dominant structure instead of defector dominant. While the ER network could not converge in 100 iterations, it still did favor cooperation. One key thing to note is that edge density also matters when it comes to convergence. When the BA and WS networks were with fewer edges, they both failed to reach convergence. However, when we increased their number of edges to match that of Facebook's they converged into a cooperator dominant state. This means that a certain number of edges is required in order for the network structure to converge and cooperation improves with edge density. However, even when convergence was reached, not all networks had the same rate of convergence. The Facebook network had the fastest rate of convergence while the ER graph could not converge within 100 iterations. We attribute convergence rate to the initial structure of the network. Because of both it's high local clustering coefficient and it's power law degree distribution, the Facebook network has the most distinct structure out of all four networks. It also had the fastest rate of convergence. The Watts Strogatz and Barabási-Albert networks were incomplete versions of the Facebook network with the Watts Strogatz network missing the power law degree distribution and the Barabási-Albert network missing a high local clustering coefficient and as a result had a slower rate of cooperator convergence. The ER graph had no small world properties and no community structure so it had the slowest rate of convergence.

**Future Directions.** For future work, we suggest the following four directions:

1. Nodes with characteristics: Nodes in networks behave in the same way (status changing and edge forming) as each other based on the specific simulation strategy used. Further work may consider having nodes with different "characteristics" together in one network. For instance, some nodes may choose to follow neighbors all the time, and other nodes tend to only care about themselves. There can also be nodes who always cooperate and nodes who always defect in the network. Such simulation would presumably provide a model that can better represent the real-life social structures.

2. Nodes with variances in behavior: For all strategies reported in the current project, cooperators always cooperate, and defectors always defect. Future work may add some noise to the nodes' behavior to model the uncertainty and miscommunications in real-life interactions.

3. Weighted network: All networks (initial networks and simulated networks) reported in this paper are undirected networks with edges showing the presence of future interaction. Another way to approach such problem would be to use weighted networks. Weights of edges between two nodes can signify the relationship between two agents in the game. Intuitively, it is more likely for nodes with edges of high weights (close friends) to decide to cooperate with each other.

4. No edge-breaking: Another variation to explore is to eliminate the edge-breaking component in the strategies. Policies with no edge breaking would force the same sets of nodes to interact with each other over and over again. Presumably, such policies would speed up convergence of the simulation due to the non-changing edge structure.

**ACKNOWLEDGMENTS.** We thank Professor Mason Porter and Abigail Hickok for providing insights on the results presented in this paper.

# Appendices

## Parameter choices

In the simulations, size of networks $N$ is either 34 or 1000 to match the size of the real networks (Karate network and Facebook network). We chose number of iterations $K = 200$ for networks of size $N = 34$, and $K = 100$ for networks with number of nodes $N = 1000$, for the sake of time and computational power. We chose the initial proportion of cooperators $r_0 = 0.5$ to investigate the network behavior based on unbiased initial conditions. We used the original prisoner's dilemma payoff matrix $M$ as indicated in Table 2. The hyperparameter for defection score calculation was chosen to be $Q_0 = 0.1$. The probability for edges to be broken when having at least one party to be a defector is $m_0 = 0.1$.

**Table 2. Payoff Matrix**

| strategy  | cooperate | defect |
|-----------|-----------|--------|
| cooperate | -1, -1    | -3, 0  |
| defect    | 0, -3     | -2, -2 |

## Code and results

The code we used can be found here on our Simulation Source Code. For a full list of simulation results, please refer to the Google Drive link